\documentstyle[preprint,aps,prl]{revtex}
\tighten
\draft
\begin{document}
\preprint{\today}
\title{Scaling behavior in the $\beta$-relaxation regime of a supercooled
Lennard-Jones mixture}
\author{Walter Kob\cite{wkob}}
\address{Institut f\"ur Physik, Johannes Gutenberg-Universit\"at,
Staudinger Weg 7, D-55099 Mainz, Germany}
\author{Hans C. Andersen\cite{hca}}
\address{Department of Chemistry, Stanford University, Stanford,
California 94305}
\maketitle

\begin{abstract}
We report the results of a molecular dynamics simulation of a
supercooled binary Lennard-Jones mixture. By plotting the self
intermediate scattering functions vs. rescaled time, we find a master
curve in the $\beta$-relaxation regime. This master curve can be fitted
well by a power-law for almost three decades in rescaled time
and the scaling time, or relaxation time, has a
power-law dependence on temperature.  Thus the predictions of
mode-coupling-theory on the existence of a von Schweidler law
are found to hold for this system; moreover, the exponents in
these two power-laws are very close to satisfying the exponent
relationship predicted by the mode-coupling-theory.  At low
temperatures, the diffusion constants also show a power-law
behavior with the same critical temperature.  However, the
exponent for diffusion differs from that of the relaxation time,
a result that is in disagreement with the theory.
\end{abstract}

\narrowtext


In the last few years a remarkable amount of activity has taken place
in the field of the glass transition and the dynamics of supercooled
liquids.  This activity was mainly spawned by the development of
sophisticated theories and novel experimental techniques with which the
predictions of these theories could be tested.  The most outstanding
example of these theories is the so-called mode-coupling-theory (MCT)
which was proposed by Bengtzelius, G\"otze and Sj\"olander and,
independently, by Leutheusser \cite{bgsleuth84}.  The central point of
MCT is the prediction of the existence of a {\it dynamic} singularity
when the temperature of a supercooled liquid is lowered below a certain
critical temperature $T_{c}$. This point of view is in contrast to
other theories that try to make a connection between the sluggish
dynamics of glass forming materials in the vicinity of the glass
transition and an underlying {\it thermodynamic} phase transition.

 For temperatures close to $T_{c}$, MCT makes detailed predictions
about the behavior of all correlation functions $<X(0)Y(t)>$ whose
dynamical variables $X$ and $Y$ have a nonvanishing overlap with the
density fluctuation $\rho_{\bf q}$ for any wave vector ${\bf q}$.
These predictions have been verified in several experiments and
computer simulations of supercooled liquids on a qualitative basis
and, more recently, also on a quantitative basis. However, there are
also experiments and computer simulations which show that the
predictions of the theory are not always
fulfilled\cite{MCTcontra,kob93b}.  The reader can find a good
introduction to the theory in the review articles by G\"otze and
G\"otze and Sj\"ogren \cite{bibles} and most of the relevant references
to the experiments and simulations in references \cite{kob93b,bibles}.

 Since experiments and computer simulations have given mixed
results as far as the validity of MCT is concerned, more tests are
clearly necessary. Moreover, {\em quantitative} tests of the theory
would be useful in order to understand whether MCT merely suggests a
method for performing scaling analyses of experimental data or whether
it is also a correct and accurate theory of the exponents and exponent
relations. Since MCT was originally developed for the description of
{\it simple} supercooled liquids we have decided to test in a
quantitative way the validity and applicability of the prediction of
MCT for such a system and this letter reports some of our findings.

 Some of the most important predictions of MCT deal with the behavior
of the correlation functions in the so-called $\beta$-relaxation
regime.  At low temperatures the time dependence of correlation
functions of supercooled liquids show a broad shoulder or even a
plateau when plotted versus the logarithm of time.  The approach to and
the subsequent departure from this plateau defines the
$\beta$-relaxation regime\cite{betafoot}. One of the main predictions
of MCT for the behavior of the correlation functions in this
$\beta$-relaxation regime is the existence of a von Schweidler law.
This law states that a correlation function $\phi(t)$ can be
written in the form
\begin{equation}
\phi(t)=f-A\left(t/\tau(T)\right)^{b}\quad .
\label{eq1}
\end{equation}
Here $f$ is the height of the plateau (the offset), often also called
the nonergodicity parameter, and $A$ and $b$ are positive constants.
$A$ and $f$ are predicted to be smooth functions of temperature if
$T>T_{c}$ and to be dependent on the type of correlation function
investigated but the exponent $b$ should be the same for all
correlation functions. Furthermore the theory predicts that the
relaxation-time $\tau(T)$ shows a power-law divergence at $T=T_{c}$ with
an exponent $\gamma$ which is related to the value of $b$.
Although the predictions of MCT concerning the von Schweidler law
have experimentally been found to be true for some systems it has so
far not convincingly been observed in computer simulations of {\it
simple} liquids. However, for a lattice gas, such a behavior has been
observed\cite{kob93a}. But since a simulation of a different lattice
gas has shown remarkable {\it disagreement} with the predictions of the
theory \cite{kob93b}, the application of the results of MCT to these
kind of systems may be problematic.  Therefore we decided to perform a
large scale computer simulation of a simple liquid in order to
investigate, among other things, the dynamics of supercooled liquids in
the $\beta$-relaxation regime.

 Molecular dynamics computer simulations are particularly well suited
for testing the predictions of MCT since they allow the measurement of
many different types of correlation functions. This freedom allows the
performance of more stringent tests of the theory than would be
possible for experiments.  In addition, in computer simulations the
measurement of the correlation functions is done in a very direct way.
This is in contrast to experiments, where in most cases a considerable
amount of interpretation and manipulation of the raw data has to be
done in order to obtain the desired quantities.  The main problems with
molecular dynamics simulations are the limited size of the systems one
can study and the time span that a simulation can cover. Since MCT is a
theory of equilibrium it is of utmost importance to make sure that the
simulated system is equilibrated at all temperatures investigated. A
very recent simulation of a glassy system has shown that failure to
equilibrate the system leads to a completely different relaxation
behavior\cite{basch94}. In order to overcome this problem we have
performed a simulation that covers a time range that is more than an
order of magnitude in time larger than previous molecular dynamics
simulations that have studied relaxation in supercooled liquids.

 The system we are dealing with in this work is a binary mixture of
classical particles. Both types of particles (A and B) have the same
mass $m$ and all particles interact by means of a Lennard-Jones
potential, i.e. $V(r)=4\epsilon ((\sigma/r)^{12}-(\sigma /r)^{6})$.
The parameters $\epsilon$ and $\sigma$ of the various interaction
potentials were chosen as follows:  $\epsilon_{AA}=1.0$,
$\sigma_{AA}=1.0$, $\epsilon_{AB}=1.5$, $\sigma_{AB}=0.8$,
$\epsilon_{BB}=0.5$, and $\sigma_{BB}=0.88$. These potentials were
truncated and shifted at a cutoff distance of $2.5\sigma$. The number
of A particles was 800, and the number of B particles was 200. These
potentials are similar to the ones used by Stillinger and Weber for
their simulation of amorphous Ni$_{80}$P$_{20}$\cite{stillweber}.  In
the following, all quantities will be expressed in a system of units in
which the unit of length is $\sigma_{AA}$, the unit of time is
$(m\sigma_{AA}^2/48\epsilon_{AA})^{1/2}$, and the unit of energy is
$\epsilon_{AA}$. The equations of motion were integrated with the
velocity form of the Verlet algorithm with a step size of 0.01 and 0.02
at high and low temperatures respectively. This step size was small
enough to restrict the fluctuations of the total energy below an
acceptable level. The system was equilibrated at high temperatures and,
by coupling it to a heat bath, subsequently slowly cooled down to low
temperatures. At each temperature the system was equilibrated for a
time which was equal to or longer than the time it took all the
correlation functions investigated to decay to zero to within the
noise. After equilibration, a long molecular dynamics calculation was
performed for the calculation of the diffusion constant and the
correlation functions.  The length of this run was also equal to or longer
than the time for the correlation functions to decay to zero, to within the
noise of the calculation.  The longest equilibrium runs were those for
the lowest temperature and had a duration of 100,000 time units, which
would correspond for an atomic liquid to a real time of about 10 ns.
This is about an order of magnitude longer than previously reported
comparable simulations.  To improve the statistics further we
determined the correlation functions for at least eight different
starting positions in phase space. More details on the simulation will
be presented elsewhere \cite{kob94}.

 In the following we will mainly concentrate on the investigation of
the dynamics in the $\beta$-relaxation regime. Since the presence of
activated, or hopping, processes will modify the prediction of the
simple version of MCT (in which these effects are neglected)
\cite{bibles} one has to determine the importance of these processes
for the system under investigation. By studying the self and distinct
part of the van Hove correlation function we have found that on the
time scale of the $\beta$-relaxation there is no secondary peak in the
self part and no peak at $r=0$ in the distinct part\cite{roux89}. Thus
we conclude that hopping processes are not relevant on the time scale
of the $\beta$-relaxation and can therefore be neglected in the
analysis of the data \cite{kob94}.  Thus the comparison of our results
with the simple version of the theory is justified.

 The space Fourier transform of the self part of the van Hove
correlation function gives the self intermediate scattering function
$F_{s}(q,t)$\cite{hansenmcdon}.  Figure~\ref{fig1} shows the time
dependence of $F_{s}(q,t)$ for the AA correlation function for all
temperatures investigated. The value of $q$ is $q_{max}$, the location
of the first maximum in the structure factor $S_{AA}(q)$ for the AA
correlation function.  The following observations can be made: 1) For
all temperatures investigated the correlation functions decay to zero
to within the noise of the data. This is strong evidence that the
runs were long enough to equilibrate the system. 2) For
high temperatures (curves to the left) the correlation functions decay
quickly to zero. When the temperature is lowered the formation of a
shoulder at intermediate times (on a logarithmic time scale) can be
observed. For the lowest temperatures (curves to the right) this
shoulder forms almost a plateau and we can clearly observe the two step
relaxation behavior observed for strongly supercooled liquids. 3) For
times around 15 time units the correlation functions for low
temperatures show a small bump.  Lewis and Wahnstr\"om recently
reported a similar observation in a computer simulation of the glass
forming liquid {\it ortho}-terphenyl\cite{lewwahn}.  These authors came
to the conclusion that the location of this bump is a finite size
effect. Therefore, and also since we have not observed this effect for
the B particles, we won't discuss this feature here any longer.

 In order to test for the presence of a scaling behavior we plotted the
data from Fig.~\ref{fig1} versus a rescaled time $t/\tau(T)$. The
value of the scaling time $\tau(T)$ was chosen at each temperature
such that $F_{s}(q,\tau)=e^{-1}$.  In Fig.~\ref{fig2} we show
the resulting plot. We can clearly recognize the presence of a master
curve. This master curve can be fitted very well by a power law of the
form given by Eq.~(\ref{eq1}) with $f=0.783$ and $b=0.488\pm0.015$.
This fit is included in the plot as well. Since the fit is valid for
almost three orders of magnitude in rescaled time it is definitely
significant. Thus we find a power law with an exponent $b$ and an
offset $f$ that are {\it independent} of temperature. This is exactly
the behavior predicted by MCT for the behavior of the correlation
function in the later part of the $\beta$-relaxation region, i.e. a
von Schweidler law with a nonergodicity parameter $f$ and an exponent
$b$ that are independent of temperature.

 In order to test whether the power-law observed in the
$\beta$-relaxation region is just the short time expansion of a
Kohlrausch-Williams-Watt (KWW) law,
$\phi(t)=f\exp(-(t/\tau)^{\beta})$, which is known to be often an
excellent approximation for the long time behavior of correlators in
glassy materials, we fitted the {\em long} time behavior of our
correlation functions with such a functional form. We found that such a
fit is very good but that the extrapolation of the fit to intermediate
times, i.e.  to the $\beta$-relaxation regime, is not good at all since
the fit falls below the data. Thus we can conclude that the power-law
observed here is not just the short time behavior of the KWW-law.

 MCT predicts that the scaling time $\tau$ in Equ.~\ref{eq1}
scales with temperature as $\tau\propto (T-T_{c})^{-\gamma}$. We
fitted the relaxation time with such a power-law and found, for the A
{\em and} B particles, for $T_{c}$ the value 0.435.  The exponent
$\gamma$ is 2.5 and 2.6 for the A and B particles respectively.
MCT also predicts a relationship between the exponents $b$ and
$\gamma$.  The value found above for $b$, when combined with
this relationship, predicts that $\gamma=2.7$, which is very
close to what we actually found.  Thus the exponent
relationship of MCT is confirmed to within the precision with
which we can obtain the exponents.
In Fig.~\ref{fig3}
we plot $\tau^{-1}$ versus $T-T_{c}$ (dotted curves).
{}From this plot we recognize that the power-law behavior is observed for
$0.466\leq T\leq0.6$. Although this range is not that large we
recognize from Fig.~\ref{fig2} that it is only for this temperature
range that the curves follow the master curve. Thus from the point of
view of MCT we have a consistent picture with respect to this range.
We also made a fit to the data with a Vogel-Fulcher law and found
that this functional form gave a good fit to the data over a
temperature range which is a bit larger than the one where the
power-law is observed\cite{kob94}. However, this observation is not
in contradiction with MCT since the main point is that the theory
works where it is supposed to work, namely close to $T_{c}$.\par

 MCT predicts that the von Schweidler law should be present not only
for $q=q_{max}$ but all values of $q$ and that the exponent $b$ should
be {\it independent} of $q$. We tested this prediction by computing the
self intermediate scattering function $F_{s}(q,t)$ for different values
of $q$ in the range $6.5\sigma_{AA}^{-1}\leq q \leq
9.6\sigma_{AA}^{-1}$, i.e. from $q$ values a bit less than $q_{max}$,
the location of the peak in the structure factor $S_{AA}(q)$, to values
up to the first minimum of $S_{AA}(q)$. In Fig.~\ref{fig4} we plot these
correlation functions as a function of $t^{b}$ with $b=0.488$. The part
of the curves that are straight lines are power laws with exponent
$b$.  We recognize that this is the case for $t^{b}$ lying
between 2-3 and about 60. This corresponds to a time interval of 4-9
time units to 3600 time units.  Thus we find that also this prediction
of MCT holds for this system.

 We have also done similar calculations for the B particles and found a
similar behavior as reported here for the A particles. The von
Schweidler exponent $b$ for the B particles is about $0.445\pm0.015$,
and thus quite close to the one of the A particles.  Since MCT predicts
these two exponents to be the same this observation is also in
accordance with the theory. Note that this kind of universality, i.e.
the presence of a von Schweidler law for all types of correlators and
that the von Schweidler exponent is independent of, or only weakly
dependent upon, the correlator, is one of the main predictions of the
theory.

 So far we have dealt only with the behavior of the correlators in the
$\beta$-relaxation regime. But MCT also proposes an intimate connection
between this regime and the relaxation on the longest time scale, the
so-called $\alpha$-relaxation regime. For example, the theory predicts
that the constant of diffusion $D$ should show a power-law divergence
at $T_c$ with an exponent $\gamma$ that is the same as the one for the
relaxation times. We have determined $D$ by fitting a straight line to
the long time asymptote of the mean squared displacement. We have
fitted $D$ with a power-law of the form $D\propto (T-T_{c})^{\gamma}$
and for {\em both} types of particles found the critical
temperature $T_{c}$ to be 0.435. The exponent $\gamma$ was 2.0 for the A
particles and 1.7 for the B particles.  These results are shown in
Fig.~\ref{fig3}, where we plot $D$ versus $T-T_{c}$ in a double
logarithmic way (dashed curves). Thus, in accordance with MCT, we find
that the critical temperature for the constant of diffusion is the same
as the one we found for the relaxation time $\tau$.
However, the $\gamma$ exponents are slightly different for the
diffusion constants of the two species, and both values are
significantly different from the $\gamma$ obtained from the
scaling time.  These findings are in contradiction to the
predictions of MCT.

 In the range were the data shows a power law behavior (i.e.  $T\leq
1.0$) we also tried to fit it with a Vogel-Fulcher law, i.e. $D\propto
\exp(-E/(T-T_{0}))$ and found  that this type of fit is clearly
inferior to the one with the power-law.

In summary, the predictions of MCT concerning the von
Schweidler law are fulfilled in quite an impressive way for the
system investigated here, but the relationship predicted by MCT
between the temperature dependence of the von Schweidler
relaxation time and the temperature dependence of diffusion do
not hold. For the $\beta$-relaxation
regime we have presented only the results of the simulation that deal
with the {\it departure} of the correlation functions from the
plateau.  But to make a {\em stringent} test of the theory it is
clearly necessary to check not only a few predictions of the theory
but as many as possible. We have done this and the results will be
presented in reference \cite{kob94}.  These results seem to
indicate\cite{kob94} that the behavior of the correlation functions in
the first part of the $\beta$-relaxation, i.e. the critical decay, is
not as well described by the theory as the second part which is
described in this letter.  Nevertheless, considering the approximations
that have to be made in order to derive the statements of the theory,
the accordance of its predictions with the results presented here is
most remarkable.

Acknowledgments: We thank J. Baschnagel for many useful discussions.
Part of this work was supported by National Science Foundation grant
CHE89-18841. We made use of computer resources provided under NSF grant
CHE88-21737.

\begin{figure}
\caption{Self intermediate scattering function $F_{s}(q,t)$ for A
particles vs. $t$. $q=7.251\sigma_{AA}^{-1}$, the location of the first
peak in the structure factor $S_{AA}(q)$. Temperatures (from left to
right): 5.0, 4.0, 3.0, 2.0, 1.0, 0.8, 0.6,0.55, 0.5, 0.475, 0.466.
\label{fig1}}

\caption{Self intermediate scattering function for the A particles for
$q=7.251\sigma_{AA}^{-1}$ vs. rescaled time $t/\tau(T)$ (solid lines).
Dashed curve: Fit with von Schweidler law. Temperatures (from right to
left) as in Fig.~2.\label{fig2}}

\caption{Inverse relaxation time (dashed curves, $T_{c}=0.430$) and
self-diffusion constant with error bars (dotted curves, $T_{c}=0.435$)
for A (circles) and B (squares) particles vs.  $T-T{c}$.  Also shown
are the fits with a power law (solid lines).  \label{fig3}}

\caption{Self intermediate scattering function for A particles for
$6.5\sigma_{AA}^{-1} \leq q \leq 9.6\sigma_{AA}^{-1}$ (from top to
bottom) vs. $t^{b}$ with von Schweidler exponent $b=0.488$.
$T=0.466$.\label{fig4}}

\end{figure}

\begin{references}
\bibitem[*]{wkob}
Electronic mail: kob@moses.physik.uni-mainz.de
%
\bibitem[\dag]{hca}
Electronic mail: fb.hca@forsythe.stanford.edu
%
\bibitem{bgsleuth84}
U. Bengtzelius, W. G\"otze, and A. Sj\"olander, J. Phys. C {\bf 17}, 5915
(1984); E. Leutheusser, Phys. Rev. A {\bf 29}, 2765 (1984).
%
\bibitem{MCTcontra}
See e.g. S. R. Nagel p. 259 in {\it Phase Transitions and Relaxation In
Systems with Competing Energy Scales} eds. T. Riste and D. Sherrington,
(Kluwer Academic Publisher, Dordrecht, 1993); B. Kim and G. F.
Mazenko, Phys. Rev. A {\bf 45}, 2393 (1992); F. Mezei, J. Non-Cryst.
Solids {\bf 130}, 317 (1991).
%
\bibitem{kob93b}
W. Kob and H. C. Andersen, Phys. Rev. E {\bf 48}, 4364 (1993).
%
\bibitem{bibles}
W. G\"otze, p. 287 in {\it Liquids, Freezing and the Glass Transition}
eds. J. P.  Hansen, D. Levesque, and J. Zinn-Justin, Les Houches.
Session LI, 1989, (North-Holland, Amsterdam, 1991); W. G\"otze and L.
Sj\"ogren, Rep. Prog. Phys. {\bf 55}, 241 (1992).
%
\bibitem{betafoot}
The $\beta$-relaxation regime of MCT should not be confused with the
idea of $\beta$-relaxation as described by Johari and Goldstein in,
e.g., G. P. Johari and M. Goldstein, J. Chem. Phys. {\bf 53}, 2372
(1970). The latter, when present in a system, generates what is
referred to as a ``$\beta$-peak'' in the literature of MCT.
%
\bibitem{kob93a}
W. Kob and H. C. Andersen, Phys. Rev. E {\bf 47}, 3281 (1993).
%
\bibitem{basch94}
J. Baschnagel, Phys. Rev. B {\bf 49}, 135 (1994).
%
\bibitem{stillweber}
T. A. Weber and F. H. Stillinger, Phys. Rev. B {\bf 31}, 1954 (1985).
%
\bibitem{kob94}
W. Kob and H. C. Andersen (preprint).
%
\bibitem{roux89}
J. N. Roux, J. L. Barrat and J.-P. Hansen, J. Phys.: Condens. Matter
{\bf 1}, 7171 (1989).
%
\bibitem{hansenmcdon}
J.-P. Hansen and I. R. McDonald, {\it Theory of Simple Liquids}
(Academic, London, 1986).
%
\bibitem{lewwahn}
L. J. Lewis and G. Wahnstr\"om, (to be published).
%
\end{references}
\end{document}